\documentclass[aps, prl, twocolumn, superscriptaddress]{revtex4}
\usepackage{graphicx}
\usepackage{dcolumn}
\usepackage{bm}
\usepackage{natbib}

\newcommand*{\eIIa}{$\tilde e \| a$}
\newcommand*{\eIIb}{$\tilde e \| b$}
\newcommand*{\eIIc}{$\tilde e \| c$}
\newcommand*{\hIIa}{$\tilde h \| a$}
\newcommand*{\hIIc}{$\tilde h \| c$}

\newcommand*{\BIIb}{$B \| b$}
\newcommand*{\cm}[1]{#1~cm$^{-1}$}

\begin{document}

\title{Evidence of electro-active excitation of the spin cycloid in TbMnO$_3$}

\author{A.~M.~Shuvaev}
\affiliation{Experimentelle Physik 4, Universit\"at W\"urzburg,
D-97074 W\"urzburg, Germany}
\author{V.~D.~Travkin}
\author{V.~Yu.~Ivanov}
\author{A.~A.~Mukhin}
\affiliation{General Physics Institute, Russian Academy of Science,
119991 Moscow, Russia}
\author{A.~Pimenov}
\affiliation{Experimentelle Physik 4, Universit\"at W\"urzburg,
D-97074 W\"urzburg, Germany}

\date{\today}

\begin{abstract}
Terahertz electromagnetic excitations in the multiferroic TbMnO$_3$
at the field-induced magnetic transition are investigated for
different orientations of the magnetic cycloid. In addition to the
electromagnon along the \textit{a}-axis, the detailed polarization
analysis of the experimental spectra suggests the existence of an
electro-active excitation for ac electric fields along the
crystallographic \textit{c}-axis. This excitation is possibly the
electro-active eigenmode of the spin cycloid in TbMnO$_3$, which has
been predicted within the inverse Dzyaloshinskii-Moriya mechanism of
magnetoelectric coupling.
\end{abstract}


\maketitle

Multiferroics represent an intriguing class of materials in which
magnetism and electricity are strongly coupled. This magnetoelectric
coupling leads to a mutual influence of magnetic and electric
ordering, which results in a rich new physics and in various
possible applications~\cite{fiebig_jpd_2005, tokura_science_2006,
khomskii_jmmm_2006, ramesh_nmat_2007}. Recently, a new class of
multiferroics has attracted enormous interest because the
ferroelectric polarization in these systems is induced by a
cycloidal ordering of the magnetic moments~\cite{kimura_nature_2003,
cheong_nmat_2007, kimura_arms_2007}. The magnetic and electric order
in these materials are accompanied by coupled spin-lattice
vibrations~\cite{baryakhtar_ptt_1970} showing strong electric dipole
activity and termed electromagnons~\cite{pimenov_jpcm_2008,
sushkov_jpcm_2008}. Initially detected in GdMnO$_3$ and
TbMnO$_3$~\cite{pimenov_nphys_2006}, the electromagnons appear to
exist in cycloidal magnetic phases of many different
multiferroics~\cite{sushkov_prl_2007, aguilar_prb_2007,
pimenov_prb_2008, kida_prb_2008, takahashi_prl_2008, kida_jpsj_2008,
lee_prb_2009, takahashi_prb_2009}. In most cases, distinct structure
of these excitations is seen including two or more separate modes.
For example, in TbMnO$_3$ one can observe at least 3 excitations:
two with the energy around $\simeq 2$~meV (\cm{10-20}) and the third
one at $\simeq 8$~meV (\cm{60}).

Ferroelectric polarization in spiral magnets has been successfully
explained on the basis of the inverse Dzyaloshinskii-Moriya (DM)
coupling between the magnetic moments~\cite{katsura_prl_2005,
sergienko_prb_2006, mostovoy_prl_2006}. Within this model a spiral
spin structure lowers the symmetry of the system leading to an
effective magnetoelectric coupling term and an electrical
polarization $P$ in the form
\begin{equation}\label{em}
    P \sim e_{ij}\times S_i \times S_j \qquad .
\end{equation}
Here $S_i$ and $S_j$ are the neighboring spins and $e_{ij}$ is the
lattice vector connecting them. Consequently, it was reasonable to
assume the same mechanism to explain the existence of the
electromagnons~\cite{katsura_prl_2007}. In this model the
electromagnons are the eigenmodes of the spin cycloid which can be
excited by electric component of the electromagnetic
wave~\cite{pimenov_jpcm_2008}. However, after initial success of
this explanation a number of experimental results contradicted the
predictions of the model. The basic example can be given by
TbMnO$_3$ at low temperatures: the DM model predicts that the
excitation conditions for the electromagnon should be tied to the
magnetic cycloid, i.e. for the bc-plane oriented cycloid the
electric polarization must be parallel to the c-axis ($P\|c$) and
the electromagnons should be excited for ac electric fields parallel
to the a-axis (\eIIa). Accordingly, for the ab-plane oriented
cycloid the polarization $P\|a$ and electromagnons for \eIIc{}
should be observed. This basically corresponds to the interchange of
the $a$ and $c$-axes in the excitation conditions. Although the
orientation of the static electric polarization has been confirmed
in many experiments~\cite{kimura_nature_2003, kimura_prb_2005,
aliouane_prb_2006}, the excitation conditions for the electromagnon
contradicted the predictions of the model: the selection rules for
the excitations remained the same (i.e. \eIIa{}) independently of
the orientation of the cycloid~\cite{pimenov_nphys_2006,
sushkov_prl_2007, aguilar_prb_2007, pimenov_prb_2008, kida_prb_2008,
takahashi_prl_2008, kida_jpsj_2008, lee_prb_2009,
takahashi_prb_2009}. In order to account for this contradiction, a
model based on Heisenberg exchange coupling between spins has been
proposed recently~\cite{aguilar_prl_2009, lee_prb_2009,
miyahara_condmat_2008}. According to this model, the structural
peculiarities of orthorhombic multiferroic manganites lead to the
selection rules \eIIa{} regardless of the orientation of the spin
cycloid.

The Heisenberg exchange model assigns the high energy electromagnons
observed between \cm{60} and \cm{80} to the zone edge
magnons~\cite{aguilar_prl_2009, lee_prb_2009, miyahara_condmat_2008,
stenberg_prb_2009}, but the nature of the lower energy elecromagnons
close to \cm{20} remains uncertain. It seems reasonable to recall
the statements of the DM model and to assume that the low-frequency
electromagnons are somehow related to the excitations of the spin
cycloid observed by inelastic neutron scattering
(INS)~\cite{senff_prl_2007, senff_jpcm_2008, senff_prb_2008} because
the characteristic frequencies coincide
closely~\cite{pimenov_prl_2009, senff_prl_2007}. This assumption is
supported by a recent observation of the magnetic excitation channel
for electromagnons~\cite{pimenov_prl_2009} which can be also seen as
antiferromagnetic resonances (AFMR) within certain excitation
conditions. However, in order to prove this assumption, the
predicted excitation conditions for the eigenmodes of the cycloid
should be found experimentally. Most specifically, one should expect
the excitation conditions along the \textit{c}-axis if the spin
cycloid is oriented in the \textit{ab}-plane. Such excitation
conditions were not observed up to now~\cite{aguilar_prl_2009,
pimenov_prl_2009}. A possible reason for this fact is the weakness
of the dielectric contribution of the spin modes within the DM
mechanism. In order to resolve this experimental difficulty,
TbMnO$_3$ seems to be an ideal candidate, because the magnetic
cycloid can be rotated between \textit{ab}-plane and
\textit{bc}-plane in external magnetic fields. Thus the ultimate
experiment must follow the \textit{c}-axis response at the
magnetically induced rotation of the cycloid.

In this work we have carried out detailed polarization analysis of
terahertz excitations in TbMnO$_3$. Special attention has been payed to
the selection rules of different excitations upon the rotation
of the spin cycloid from \textit{bc}- to \textit{ab}-plane in external
magnetic field \BIIb{}-axis. It was found that a new excitation
arises in the high-field phase with excitation conditions \eIIc{} and
\hIIa{}. We argue that this excitation could not be explained by
purely magnetic contribution but carries substantial electric component.

The transmittance experiments at terahertz frequencies (\cm{3} $<
\nu <$ \cm{30}) have been carried out in a Mach-Zehnder
interferometer arrangement~\cite{volkov_infrared_1985,
pimenov_prb_2005} which allows measurements of amplitude and phase
shift in a geometry with controlled polarization of radiation. The
experiments in external magnetic fields up to 8~T have been
performed in a superconducting split-coil magnet with polypropylene
windows. Single crystals of TbMnO$_3$ have been grown using the
floating-zone method with radiation heating. The samples were
characterized using X-ray, magnetic, dielectric and optical
measurements~\cite{schmidt_epjb_2009, pimenov_jpcm_2008}. The
results of these experiments including the magnetic phase diagram
are closely similar to the published results~\cite{kimura_prb_2005}.

\begin{figure}
\includegraphics[width=1.0\linewidth, clip]{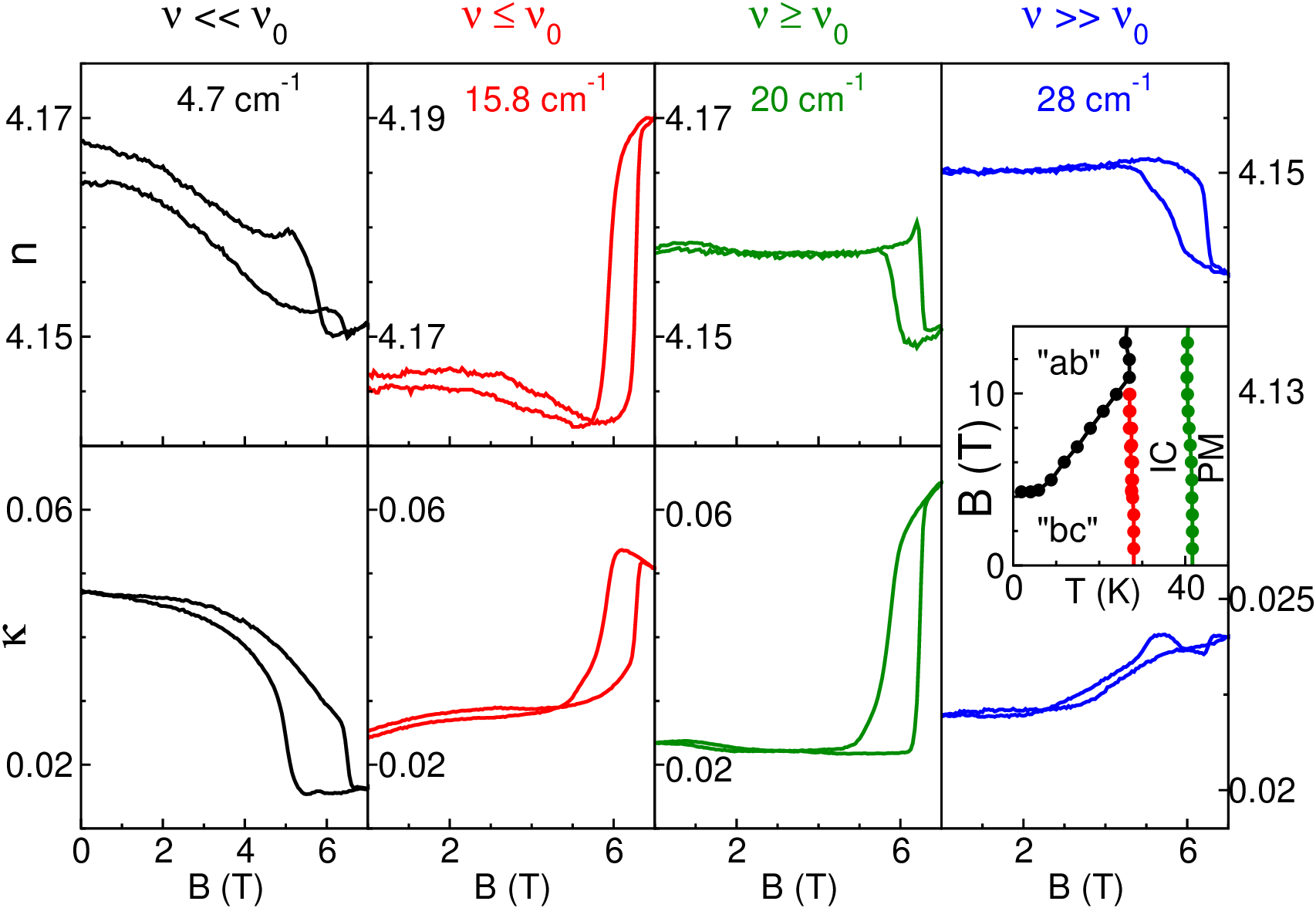}
\caption{Dependence of refractive index ($n=Re(\sqrt{\varepsilon
\mu})$, upper panels) and absorption coefficient
($\kappa=Im(\sqrt{\varepsilon \mu})$, lower panels) in TbMnO$_3$ on
magnetic field \BIIb{}-axis at various frequencies on crossing the
phase transition from the $bc$- to the $ab$-oriented magnetic
cycloid. Polarization of incident wave is \eIIc{}, \hIIa{}, where
$\tilde e$ and $\tilde h$ are electric and magnetic ac fields of the
electromagnetic wave. The inset shows $B$-$T$ phase diagram of
TbMnO$_3$ for \BIIb{}~\protect{\cite{kimura_prb_2005}}.
``\textit{ab}'' and ``\textit{bc}'' denote \textit{ab}-plane and
\textit{bc}-plane oriented cycloids, respectively, PM - paramagnetic
and IC - sinusoidal phases.} \label{scans}
\end{figure}

The paramagnetic phase in TbMnO$_3$ above 40~K is followed by a
sinusoidally-modulated antiferromagnetic state which transforms
below 28~K into the cycloidal spin structure oriented within the
\textit{bc}-plane~\cite{quezel_physica_1977, kenzelmann_prl_2005}
(inset in Fig. \ref{scans}). According to the symmetry
analysis~\cite{mostovoy_prl_2006, cheong_nmat_2007} and Eq.
(\ref{em}), static electric polarization along the \textit{c}-axis
is allowed in the low-temperature phase. In external magnetic fields
the spin-cycloid rotates from \textit{bc}-plane towards the
\textit{ab}-plane~\cite{senff_prb_2008, aliouane_prb_2006}.
Correspondingly, the electric polarization rotates from $P \|
c$-axis to the $P \| a$-axis~\cite{kimura_nature_2003,
kimura_prb_2005, aliouane_prb_2006}.

Dynamic experiments on TbMnO$_3$ in external magnetic fields reveal
that the $c$-axis properties are indeed sensitive to the orientation
of the cycloid. The examples of such changes are shown in
Fig.~\ref{scans} which represents the terahertz properties of
TbMnO$_3$ in external magnetic field \BIIb{} inducing the transition
from $bc$-plane to $ab$-plane oriented cycloid. The magnetic field
scans in these experiments were made in the geometry with \eIIc{}
and \hIIa{} and at $T = 10$~K. The data are represented as
refractive index $n + i \kappa= \sqrt{\varepsilon \mu}$ as both
electric and magnetic contributions could be mixed in this
experimental geometry. The transition from \textit{bc}-plane to
\textit{ab}-plane cycloid in the high magnetic fields is clearly
seen. Already at this point it is evident that the observed changes
are strongly frequency dependent. Here the data at \cm{4.7} is
influenced by a Tb-mode around \cm{5}~\cite{pimenov_prl_2009} which
disappears in the high-field phase with the \textit{ab}-plane
cycloid. This leads to a substantial decrease of the absorption
($\kappa$(\cm{4.7})) and reveals a bit complicated structure in
refractive index below \cm{10}. The changes observed at \cm{4.7} can
be well understood assuming a suppression of a Lorentzian mode
situated between 5 and \cm{6}.

Three higher frequency scans (15.8, 20, and \cm{28}) in
Fig.~\ref{scans} show more systematics, and can be reduced to a
growth of the absorption mode around \cm{20} in the phase with the
$ab$-plane cycloid. This is a typical behavior for a Lorentz
oscillator which appears close to \cm{20} simultaneously with the
\textit{ab}-plane cycloid. Indeed, strong additional absorption
arises near the frequency $\simeq$~\cm{20} and is substantially
reduced above and below this frequency (lower panels, \cm{28} and
\cm{16}, respectively). In addition to a strongly increased
absorption around the resonant frequency, the refractive index in
Lorentzian model may in rough approximation be simplified to the
following statements: $n(\nu) \approx n_\infty+\Delta n$ for
$\nu<\nu_0$ and $n(\nu) \approx n_\infty-\Delta n(\nu_0/\nu)^2$ for
$\nu>\nu_0$. Here $n_\infty$ is the contribution from the
high-frequency processes, $\Delta n$ reflects the strength of the
Lorentzian mode, and $\nu_0 \approx$
 \cm{20} is the resonance frequency. Therefore, after the growth of
the Lorentzian ($\Delta n = 0 \rightarrow \Delta n \neq 0$) at the
magnetic transition we expect an increase of the refractive index
below the resonance frequency and a decrease above it. This is
roughly the behavior of the refractive index, observed at \cm{15.8}
($\nu<\nu_0$) and \cm{28} ($\nu > \nu_0$). The main changes, seen in
Fig. \ref{scans}, appears at the magnetic transition and are
therefore related to the rotation of the magnetic cycloid. However,
better understanding of the underlying processes can be obtained
within a direct analysis of the spectra in the relevant frequency
range as presented below.

\begin{figure}
\includegraphics[width=0.6\linewidth, clip]{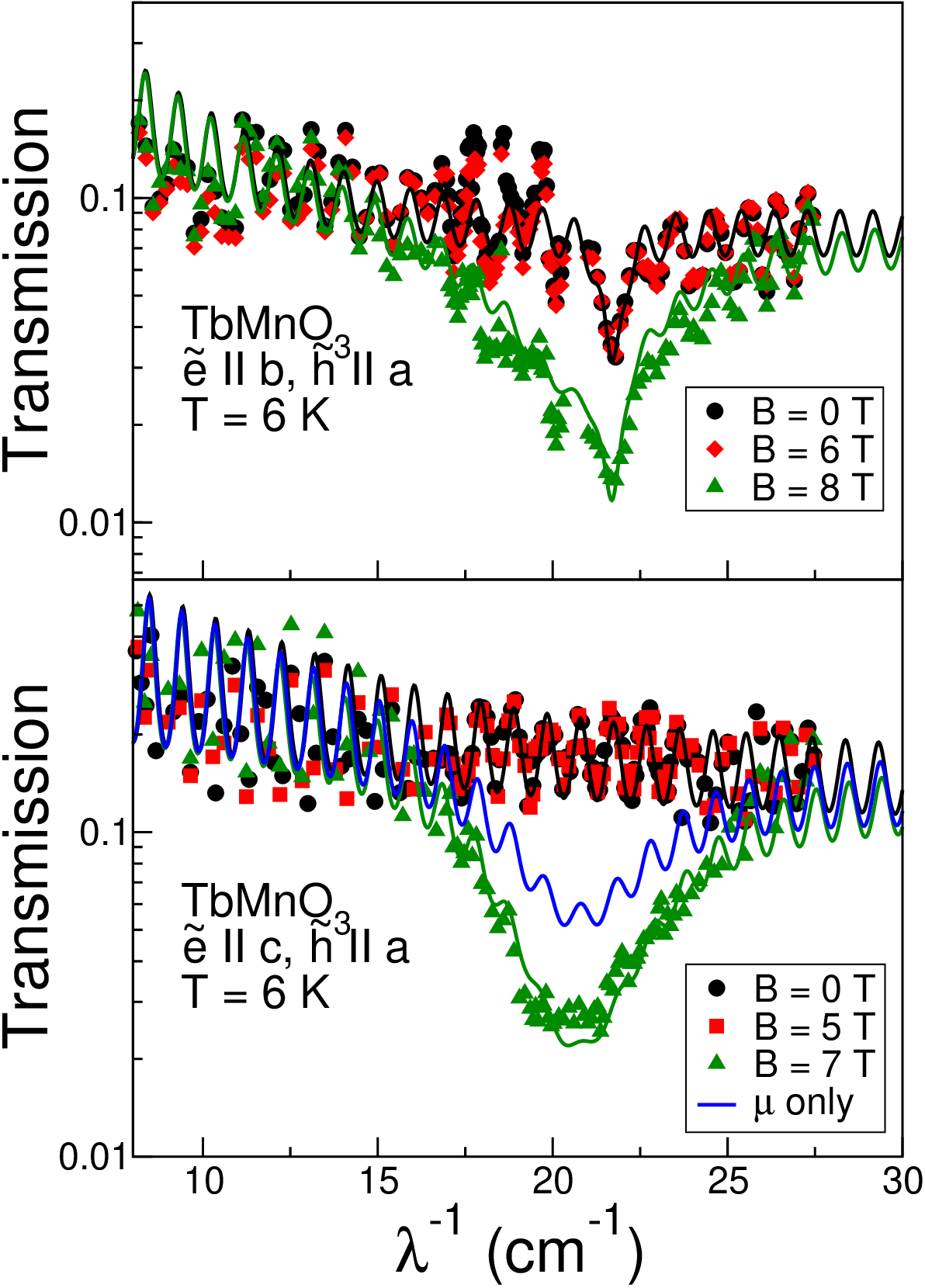}
\caption{Transmittance spectra of TbMnO$_3$ in external magnetic
fields \BIIb{} for different experimental geometries. Symbols are
experimental data and solid lines are fits with Lorentz oscillators
as discussed in the text.}
\label{spectra}
\end{figure}

Figure~\ref{spectra} shows the field dependent spectra for two
different geometries of the experiment. The thicknesses of the
samples are similar for both orientations: 1.24~mm (upper panel) and
1.33~mm (lower panel), respectively. The spectra in the lower panel
with \eIIc{} and \hIIa{} correspond well to the known
results~\cite{pimenov_prl_2009, aguilar_prl_2009} and show a mode at
about \cm{21} which appears after the field induced reorientation of
spin cycloid to the {\it ab}-plane. Based on the weakness of this
mode, both in Ref.~\cite{aguilar_prl_2009} and in
Ref.~\cite{pimenov_prl_2009} it has been concluded that the mode
around \cm{21} is of purely magnetic origin and represent an
antiferromagnetic resonance of the magnetic cycloid for \hIIa-axis.
Indeed, the strength of this mode ($\Delta \varepsilon \sim 0.05$,
assuming electric origin) is extremely weak compared to
electromagnon observed for \eIIa{} ($\Delta \varepsilon \sim
2$)~\cite{pimenov_nphys_2006, pimenov_jpcm_2008}. The mode in
Fig.~\ref{spectra} is observed for the \textit{ab}-plane cycloid and
within \hIIa{} excitation conditions. Tracing this mode back into
the \textit{bc}-oriented cycloid in zero external fields, it can be
expected to originate from the excitation conditions \hIIc{}. (This
corresponds to the interchanging of the \textit{a}-axis and
\textit{c}-axis). Indeed, an AFMR mode excited for \hIIc{} of the
similar strength has been observed around \cm{21} in the phase with
the $bc$-plane cycloid~\cite{pimenov_prl_2009}.

A careful comparison of both panels in Fig.~\ref{spectra} reveals
interesting difference between two excitation conditions.
The strength of the mode in the geometry where \eIIb{} is roughly
the half of that where \eIIc{}. This strongly suggests that for geometry
in which \eIIc{} the electric dipole contribution is indeed
measurable and represent the previously unobserved \eIIc{}
counterpart of the electromagnon. These results agree well with the
original explanation of the electromagnons as electrically active
eigenmodes of the cycloidal structure~\cite{katsura_prl_2007,
pimenov_jpcm_2008}.

In order to make the discussion quantitative, the experimental
spectra in the upper panel of Fig.~\ref{spectra} were fitted with
magnetic Lorentz oscillator. If we now take the parameters of the
mode from the geometry with \eIIb{} and plot the expected
transmittance spectra for the geometry \eIIc{} we obtain the
absorption value which is too weak compared to the experiment (the
``$\mu$ only'' curve in the lower panel of Fig.~\ref{spectra}). The
only possible explanation is that this mode has distinct non-zero
electric contribution along the \textit{c}-axis. The actual fit for
this geometry was obtained by taking parameters of the magnetic
oscillator from \eIIb{}, \hIIa{} geometry and adding an electric
oscillator with the same resonance frequency $\nu_0$ = \cm{20.7} and
line width $\gamma$ = \cm{4.9} as the magnetic one. The reasoning
behind this assumption is that both contributions are electric and
magnetic parts of the same eigenmode of the spin cycloid. The
strengths of both components is given by $\Delta \mu = 0.0038$ and
$\Delta \varepsilon = 0.05$, respectively.

\begin{figure}
\includegraphics[width=0.7\linewidth, clip]{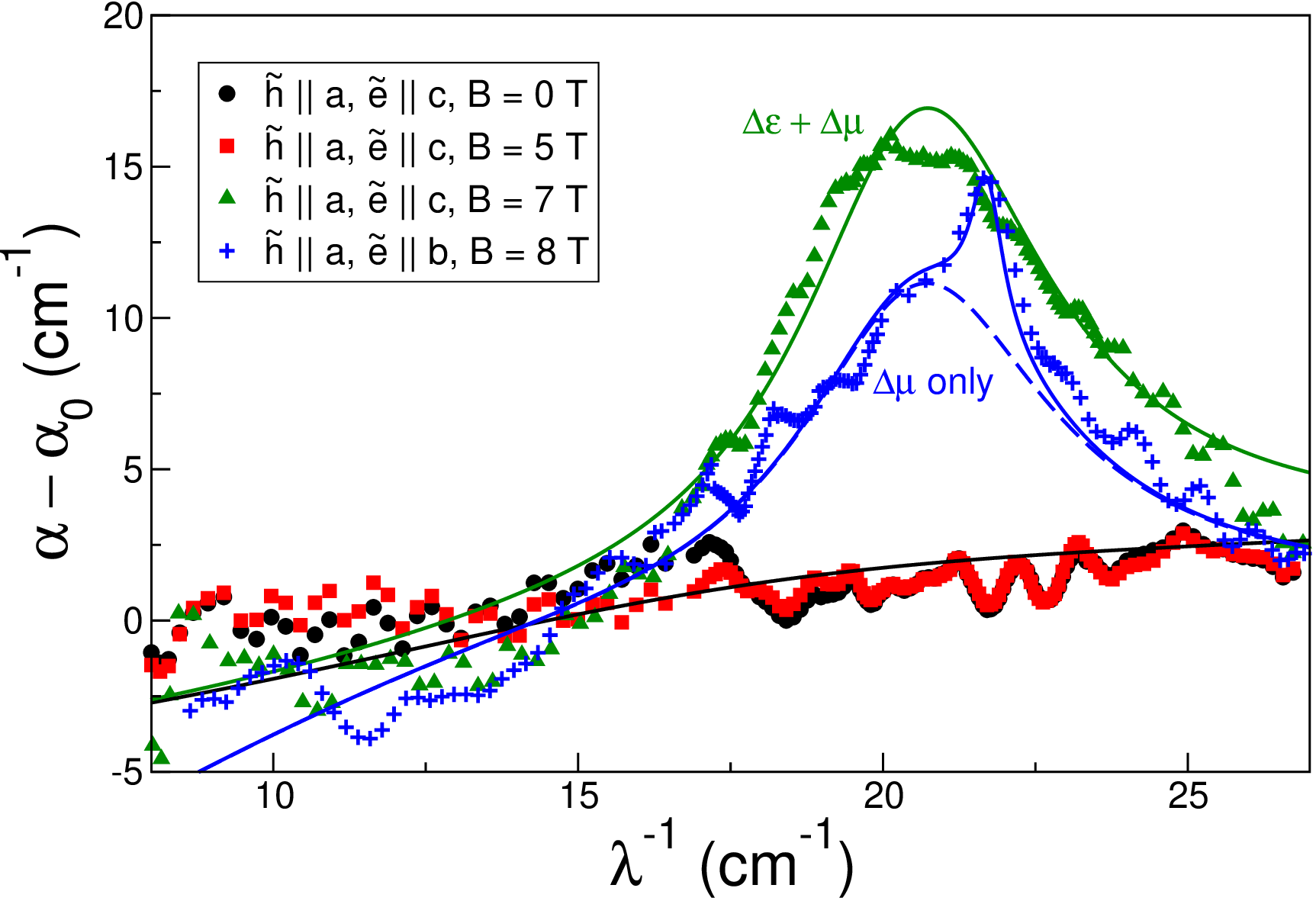}
\caption{Linear absorption coefficient $\alpha = 4 \pi \kappa \lambda^{-1}$
in TbMnO$_3$ for different polarizations showing the emergence
of an additional mode in the phase with the \textit{ab}-cycloid
(green triangles and blue crosses). Symbols are experimental data
calculated from transmittance spectra as described in the text.
Lines are model calculations using Lorentz oscillators with the same
parameters as in Fig.~\ref{spectra}. The data are shifted vertically
to eliminate background absorption in different samples.}
\label{alpha}
\end{figure}

Figure~\ref{alpha} shows the linear absorption coefficients $\alpha
= 4 \pi \kappa \lambda^{-1}$ which allow direct comparison of
different experimental geometries. Symbols were calculated from
transmittance spectra $T$ using the expression $T = (1 - R)^2 \exp(-
\alpha d)$, where $R=|(\sqrt{\varepsilon \mu}-1)/(\sqrt{\varepsilon
\mu}+1)|^2$ is the reflectance on the boundary between the air and
the sample, and $d$ is the sample thickness. Solid lines are model
calculations using the same parameters as in Fig.~\ref{spectra}. One
can see again that the mode at \cm{21} is stronger in the case of
\eIIc{} and \hIIa{} excitation. (A weak narrow mode seen close to
\cm{22} for \eIIb{}, \hIIa{} geometry is possibly due to impurities
in the sample. The strength of this mode is at least an order of
magnitude smaller than the strength of the broad mode and doesn't
change the overall picture.)

The mode intensity for the ``main'' \eIIa{} electromagnon ($\Delta
\varepsilon_a \simeq 2$~\cite{pimenov_nphys_2006}) is about 40 times
stronger than electric contribution along the \textit{c}-axis
($\Delta \varepsilon_c \simeq 0.05$) observed in the present
experiment. The large electromagnon absorption along the
\textit{a}-axis was one of the challenging questions in explaining
its origin. The relatively weak static electric polarization doesn't
fit well with the large dielectric absorption of electromagnon if
both are caused by Dzyaloshinskii-Moriya
interaction~\cite{aguilar_prl_2009}. On the contrary, the Heisenberg
exchange mechanism~\cite{aguilar_prl_2009, lee_prb_2009,
miyahara_condmat_2008} seems to explain well the intensities of at
least the high-frequency electromagnons above \cm{50}. In this model
the edge-zone magnon couples to alternating orthorhombic distortions
at oxygen sites via symmetric Heisenberg exchange interaction. This
leads to the coupling of the zone edge magnon to homogeneous
electric fields along the \textit{a}-axis. As the symmetric
interaction is much stronger than the relativistic DM coupling, the
hybridized electromagnon has enough strength to explain the
experimental intensities for \eIIa{}. Much
weaker~\cite{aguilar_prl_2009} DM component cannot be seen in this
experimental geometry because of the dominance of the intensity
induced by the Heisenberg exchange coupling. On the contrary,
rotating the magnetic cycloid towards the \textit{ab}-plane, both
contributions can be well separated experimentally. The Heisenberg
exchange part remain oriented along the \textit{a}-axis, as
confirmed by different experimental groups~\cite{kida_prb_2008,
takahashi_prb_2009, aguilar_prl_2009, pimenov_prl_2009}. The weak DM
electromagnon rotates with the cycloid and can be clearly observed
in the present experiment as electric contribution along the
\textit{c}-axis.

One remaining question is: why we observe only one mode in the
high-field phase? The probable reason is that one of two modes is
too weak and is not seen in the spectra. This argument is supported
by recent inelastic neutron scattering
experiments~\cite{senff_prb_2008}. In these experiments the modes of
the \textit{ab}-plane spin cycloid have been investigated. Although
this \textit{ab}-plane orientation has been achieved using external
magnetic field along the \textit{a}-axis, the comparison to the
present results is still very instructive. It has been observed that
the excitations of the \textit{ab}-cycloid are dominated by a strong
mode at 2.25~meV~\cite{senff_prb_2008}. This frequency corresponds
well to the excitation at \cm{21}, seen in
Figs.~\ref{spectra},\ref{alpha}.

In conclusion, we performed detailed polarization analysis of the
electric and magnetic excitations in TbMnO$_3$ in the high-field
phase where spin cycloid rotates from \textit{bc}- to
\textit{ab}-plane. The observed excitation at \cm{21} could not be
described by purely magnetic contribution as was suggested
previously. We argue that this excitation is the missing
electro-active eigenmode of the spin cycloid. The weakness of this
mode is in agreement with the Dzyaloshinskii-Moriya contribution to
the dynamical magnetoelectric coupling in TbMnO$_3$, which
represents a relativistic counterpart to the Heisenberg exchange
mechanism.

This work has been supported by DFG (PI 372) and by RFBR
(09-02-01355).

\bibliography{magnetoelectric}

\end{document}